\documentclass[12pt,a4paper]{article}

\usepackage{graphicx}
\usepackage{amssymb}
\usepackage{amsmath}
\usepackage{url}
\usepackage{color}
\usepackage[lined,boxed,commentsnumbered]{algorithm2e}
\usepackage[normalem]{ulem}

\usepackage[T1]{fontenc}

\usepackage{authblk}
\usepackage{setspace}
\title{FPGA-Based Bandwidth Selection for Kernel Density Estimation Using High Level Synthesis Approach}

\author[*]{Artur Gramacki}
\author[*]{Marek Sawerwain}
\author[**]{Jaros\l{}aw Gramacki}
\affil[*]{
Institute of Control and Computation Engineering,
University of Zielona G\'o{}ra, Poland, 
email: \texttt{\{A.Gramacki, M.Sawerwain\}@issi.uz.zgora.pl}
}
\affil[**]{
Computer Center,
University of Zielona G\'o{}ra, Poland,
email: \texttt{J.Gramacki@ck.uz.zgora.pl}
}

\begin{document}
\date{} 
\maketitle

\begin{abstract}
FPGA technology can offer significantly hi\-gher performance at much lower power consumption than is available from CPUs and GPUs in many computational problems. Unfortunately, programming for FPGA (using ha\-rdware description languages, HDL) is a difficult and not-trivial task and is not intuitive for C/C++/Java programmers. To bring the gap between programming effectiveness and difficulty the High Level Synthesis (HLS) approach is promoting by main FPGA vendors. Nowadays, time-intensive calculations are mainly performed on GPU/CPU architectures, but can also be successfully performed using HLS approach. In the paper we implement a bandwidth selection algorithm for kernel density estimation (KDE) using HLS and show techniques which were used to optimize the final FPGA implementation. We are also going to show that FPGA speedups, comparing to highly optimized CPU and GPU implementations, are quite substantial. Moreover, power consumption for FPGA devices is usually much less than typical power consumption of the present CPUs and GPUs. 

\vspace{5 mm}
\end{abstract}

\section{Introduction} \label{sec:introduction}

The probability density function (PDF) is a key concept in statistics with many practical applications, see for example \cite{Kulczycki2010}, \cite{Silverman-1986}. Constructing the most adequate PDF from the observed data is still an important and interesting research problem, especially for large datasets. PDFs are often calculated using nonparametric data-driven methods. One of the most popular nonparametric method is the kernel density estimation (KDE) \cite{Wand-1995}. However, a very serious drawback of using KDE is the large number of calculations required to compute them as well as to find the optimal bandwidth (smoothing) parameter (time complexity $O(n^2)$). 

In this paper we investigate the possibility of utilizing field-programmable gate arrays (FPGA) to accelerate finding of such the optimal bandwidth. Towards the needs of the paper we have selected one popular and often used algorithm called in literature the PLUGIN \cite{Wand-1995}. This work can be considered as a continuation and extension of the paper \cite{Gram-2013}, where the authors utilize GPUs for speeding up optimal bandwidth selection. One of the algorithm analyzed in that paper was the above mentioned PLUGIN one. 

Generally, there are two methodologies for speeding up complex numerical algorithms: software-based and ha\-rd\-w\-are-based. In this paper we concentrate only on hardware-based methods. The commonly known approaches are as follows: 
(a) computing on general purpose multicore CPU microprocessors,
(b) computing on distributed environments (e.g. clusers, grids, etc.),
(c) computing on GPUs, \cite{Steffen2010} 
(d) computing on DSP units
and 
(e) computing on FPGA chips \cite{Lei2013, Lin2013, PEDROFERLIN2011b, Taherkhani, Wyrwol2013}.

In the paper we are concerned with FPGA approach. In \cite{Fahmy2010} the author considers a problem how to use FPGA for fast computing of PDFs using direct VHDL programming approach. However, the problem we are concerning is of different nature as we concentrate our attention for computing the optimal bandwidth for PDF (see Chapter \ref{sec:KDE-Band}). 

To develop the final FPGA design we use the High Level Synthesis (HLS) approach \cite{CoussyMorawiec2008}, \cite{Matai2014}, where no direct hardware description language (HDL) coding is needed (typically in VHDL or Verilog languages\footnote{It is worth to note that OpenCL framework, which is commonly used by GPU programmers, becomes also available for FPGA devices. Nowadays, OpenCL is offered by \emph{Altera SDK for OpenCL} to easily implement OpenCL applications for FPGA. Recently, Xilinx announced a similar solution, namely \emph{SDAccel Development Environment} for OpenCL, C, and C++. 
}).   

The remainder of the paper is organized as follows. In section \ref{sec:KDE-Band} we turn our attention to give the reader some preliminary information on KDE and bandwidth selection. In section \ref{sec:plugin} we give detailed mathematical formulas for calculating optimal bandwidth using the PLUGIN method. In section \ref{sec:fpga-impl} we cover all the necessary details on our FPGA-based implementation. We also present practical experiments we carried out and discuss the results. In section \ref{sec:concluding} we conclude the paper. 

\section{Kernel Density Estimation and Bandwidth Selection}\label{sec:KDE-Band}

Let us consider a continuous \emph{univariate} random variable $X$ and let assume its probability density function $f$ exists but is unknown.  Its estimate, usually denoted by $\hat{f}$, will be determined on the basis of a random sample of size $n$, that is $X_1, X_2, ..., X_n$ (our experimental data). In such a case, a $1$-dimensional kernel density estimator $\hat{f}(x,h)$ of a real density $f(x)$ for random sample $X_1,X_2,\ldots,X_n$ is given by the following formula
\begin{equation}\label{eq:kde-def}
\hat{f}(x,h) 
= 
\frac{1}{n h}\sum_{i=1}^n K \left( \frac{x-X_i}{h} \right).
\end{equation}
$h$ is a positive real number called \emph{smoothing parameter} or \emph{bandwidth} and $K(\cdot)$ is the \emph{kernel function} -- a symmetric function that integrates to one. In practical applications $K(\cdot)$ has often the Gaussian normal form, that is
\begin{align}\label{eq:gaussian}
K(u) = \frac{1}{\sqrt{2\pi}} \exp \left( -\frac{1}{2} u^2 \right).
\end{align}
If we have the bandwidth $h$, we can determine the estimator $\hat{f}(x,h)$ of the unknown  density function $f(x)$ using  formula (\ref{eq:kde-def}). The bandwidth $h$ is the parameter which exhibits a strong influence on the resulting KDE. 

Formula (\ref{eq:kde-def}) can be easily extended to the \emph{multivariate} case. In the most general variant the scalar bandwidth $h$ is replaced by the \emph{unconstrained} bandwidth matrix $H$ (which is symmetric and positive definite). Also  (\ref{eq:gaussian}) is generalized to the multivariate case. Two commonly used kernel types are \emph{product} and \emph{radial} (also known as \emph{spherically symmetric}) ones. 

\begin{figure}[!t]
	\centering
	\includegraphics[width=13cm]{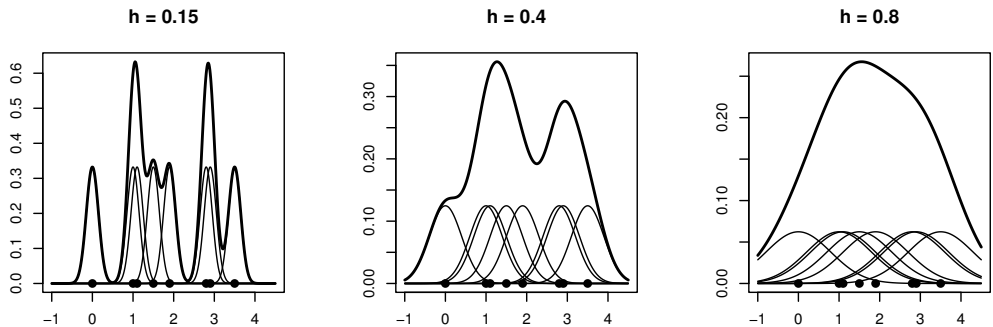}
	\caption{An example of using kernel density estimators for determining the probability density function.} 
	\label{fig:kernel-demo}
\end{figure}

As an example of how KDE works consider a toy dataset of 8 data points:
$
X_i=\{0, \, 1, \, 1.1, \, 1.5, \, 1.9, \, 2.8, \, 2.9, \,  3.5\} . 
$
Three different KDEs based on these data are depicted in Figure \ref{fig:kernel-demo}. It is easy to notice how the bandwidth $h$ influences the shape of the KDE curve. Lines in bold show the estimated PDFs, while normal lines show the shapes of individual kernel functions $K(x)$ (Gaussians). Dots represent the data points $X_i$.

Choosing the best value of $h$ is not a trivial task and this problem was and still is extensively studied in literature \cite{Chacon-2010}, \cite{Chacon-2012}. Currently available selectors can be roughly divided into 3 classes \cite{Silverman-1986}, \cite{Wand-1995}. 

The first class uses very simple and easy to calculate mathematical formulas. They were developed to cover a wide range of situations, but do not guarantee being enough close to the optimal bandwidth. They are often called rules-of-thumb methods. 

The second class contains methods based on least squares and cross-validation ideas with more precise mathematical arguments, but they require much more computational effort. However, in reward for it, we get bandwidths more accurate for a wider range of density functions. 

The third class contains methods based on plugging in estimates of some unknown quantities that appear in formulas for the asymptotically optimal bandwidth. 

One selected method from the third class is investigated in the paper. The method is briefly presented in Chapter \ref{sec:plugin} and from now on it will abbreviated as the PLUGIN.

\section{The PLUGIN Method and Data Preprocessing}\label{sec:plugin}
In Algorithm~\ref{alg:plugin} we give recipe for calculation of the optimal bandwidth using the PLUGIN method (the symbols used are exactly such as in the book \cite{Wand-1995}). All the necessary details on the method, as well as details on deriving of particular mathematical formulas can be found in many source materials, see for example books \cite{Kulczycki2005, Silverman-1986,Simonoff-1996,Wand-1995}. 

\begin{figure}
	\centering
	\includegraphics[width=10cm]{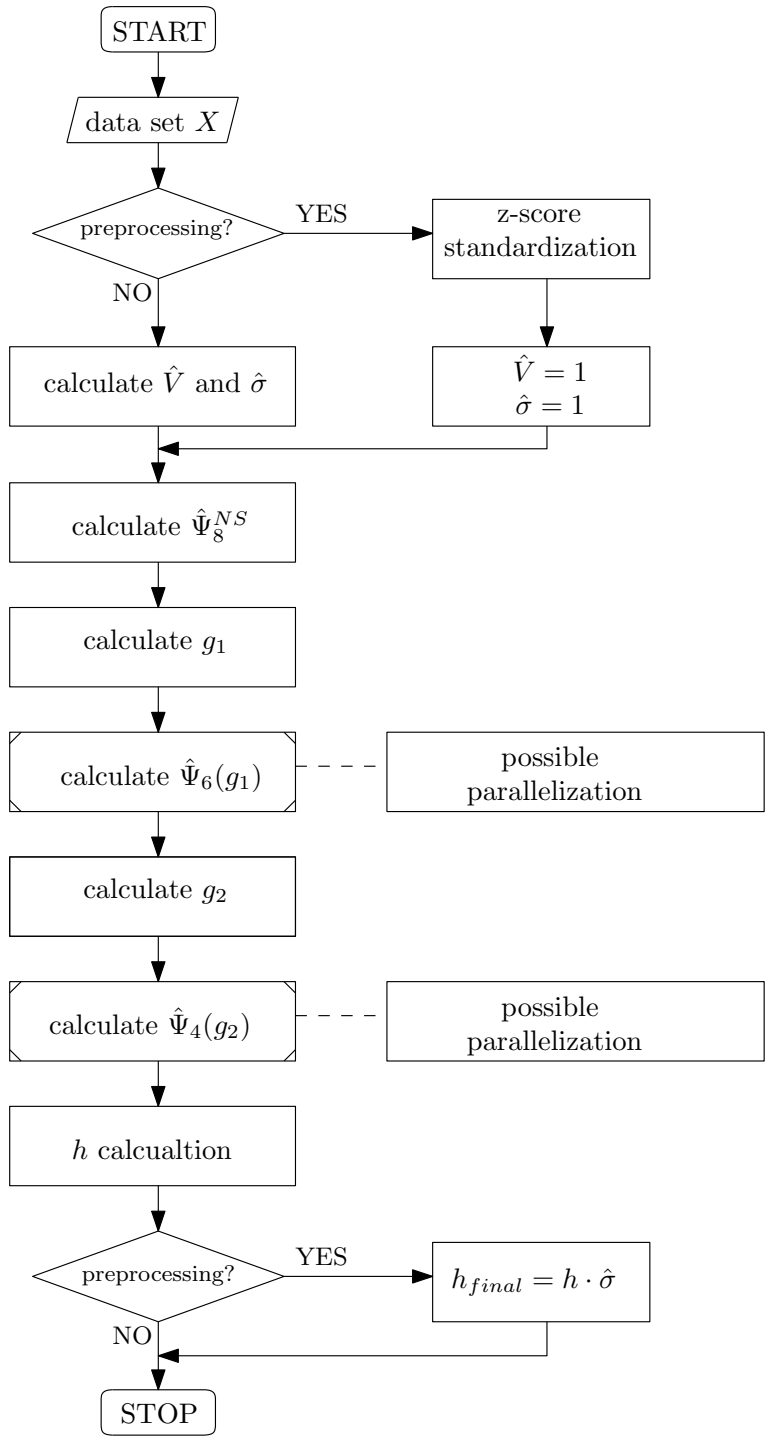}
	\caption{Flowchart of the PLUGIN algorithm with optional data preprocessing (z-score standardization)} 
	\label{flowchart:of:plugin}
\end{figure}

It is important to stress that the PLUGIN algorithm is a strictly \emph{sequential} computational process (see Figure \ref{flowchart:of:plugin}; parallel processing is possible only internally in Steps IV and VI) as every step depends on the results obtained in the previous steps. First we calculate the variance and the standard deviation estimators of the input data, see Step I in Algorithm \ref{alg:plugin}. Then we calculate some more complex formulas from Step II to Step VI. 
Finally, we can substitute them into equation given in Step VII to get the searched optimal bandwidth value $h$.
\begin{algorithm} 
\footnotesize
\KwData{data set $X$, contains $n$ elements}
\KwResult{value $h$ represents the optimal bandwidth for kernel density estimation}
\textbf{Step I:} \emph{Calculate the estimates of variance ($\hat{V}$) and standard deviation ($\hat{\sigma}$}):
\begin{displaymath} 
  \hat{V} \gets \frac{1}{n-1}\sum^{n}_{i=1}X^2_i-\frac{1}{n(n-1)}\left(\sum^n_{i=1}X_i\right)^2,  \;\;  \hat{\sigma} \gets \sqrt{\hat{V}}.
\end{displaymath} 

\textbf{Step II:} \emph{Calculate the estimate $\hat{\Psi}^{NS}_8$ of functional $\Psi_8$}:
\begin{displaymath} 
  \hat{\Psi}^{NS}_8 \gets \frac{105}{32\sqrt{\pi}\hat{\sigma}^9}.
\end{displaymath}

\textbf{Step III:} \emph{Calculate the bandwidth of the kernel estimator of function $f^{(4)}$ (4th derivative of  function $f$, that is $f^{(r)}=\frac{d^rf}{dx^r}$)}: 
\begin{displaymath} 
  g_1 \gets \left( \frac{-2K^6(0)}{\mu_2(K)\hat{\Psi}_8^{NS}n} \right)^{1/9}, \;\;
  K^6(0) = -\frac{15}{\sqrt{2\pi}}, \;\; \mu_2(K) = 1
\end{displaymath}

\textbf{Step IV:} \emph{Calculate the estimate $\hat{\Psi}_6(g_1)$ of functional $\Psi_6$}:
\begin{displaymath} 
  \begin{split}
  \hat{\Psi}_6(g_1) 
   & \gets  \frac{1}{n^2 g_1^7} \left[ \sum_{i=1}^{n}\sum_{j=1}^{n} K^{(6)} \left( \frac{X_i - X_j}{g_1} \right)\right],  \\ 
  K^{6}(x) 
  &= \frac{1}{\sqrt{2\pi}}\left( x^6 - 15x^4 +45x^2 -15 \right) e^{-\frac{1}{2}x^2}.
  \end{split}
\end{displaymath}

\textbf{Step V:} \emph{Calculate the bandwidth of the kernel estimator of function $f^{(2)}$}:
\begin{displaymath} 
  g_2 
  \gets
  \left( \frac{-2K^4(0)}{\mu_2(K)\hat{\Psi}_6(g_1)n} \right)^{1/7}, \;
  K^4(0) 
  = \frac{3}{\sqrt{2\pi}}, \;\; \mu_2(K) = 1
\end{displaymath}

\textbf{Step VI:} \emph{Calculate the estimate $\hat{\Psi}_4(g_2)$ of functional $\Psi_4$}:

  \begin{displaymath} 
  \begin{split}
  \hat{\Psi}_4(g_2) 
  & \gets 
  \frac{1}{n^2 g_2^5} \left[ \sum_{i=1}^{n}\sum_{j=1}^{n} K^{(4)} \left( \frac{X_i - X_j}{g_2}\right)\right],  \\
  K^{4}(x) 
  &= 
  \frac{1}{\sqrt{2\pi}}\left( x^4 - 6x^2 + 3 \right) e^{-\frac{1}{2}x^2}.
  \end{split}
  \end{displaymath}

\textbf{Step VII:} \emph{Calculate the final value of the bandwidth $h$}:
  \begin{displaymath} 
  h
  \gets \left( \frac{R(K)}{\mu_2(K)^2 \hat{\Psi}_4(g_2) n}\right)^{1/5}\; , \;
  R(K) = \frac{1}{2\sqrt{\pi}}, \;\; \mu_2(K) = 1
  \end{displaymath}
\caption{Main computational steps of the PLUGIN algorithm}
\label{alg:plugin}
\normalsize
\end{algorithm}

Our implementation of the Algorithm \ref{alg:plugin} is carried out in fixed-point arithmetic (see section \ref{sec:impl-prelim}). Unfortunately, using the raw data while conducting the required calculations, threatens a potential problems with overflow, especially while calculating the value of
$\hat{\Psi}^{NS}_8$, see Step II in Algorithm~\ref{alg:plugin}. Note that the estimate of standard deviation in $\hat{\Psi}^{NS}_8$ is raised to the power of $9$. For large values of $\sigma$ it results in extremely small values of $\hat{\Psi}^{NS}_8$. The above problems can be successfully overcome if the input datasets are \emph{standardized} using the \emph{z-score} formula, that is 
\begin{equation}\label{eq:z-score}
Z_i=\frac{X_i-\mu}{\sigma}
\end{equation}
where $\mu$ and $\sigma$ are mean and standard deviation of the original vector $X$ respectively. \emph{Z-score} guarantees that $ \hat\sigma=1$ in $\hat{\Psi}^{NS}_8$ and, consequently,  $\hat{\Psi}^{NS}_8$ entity has simply a constant value. 

Applying the data standardization requires an extra operation on the $h$ value in Step VII in Algorithm~\ref{alg:plugin}, that is 
\begin{equation}\label{eq:h-final-z-score}
h_{\mathrm{final}} = h \cdot \hat{\sigma}.
\end{equation}
where $h$ is the bandwidth calculated for the standardized dataset and $\hat{\sigma}$ is the standard deviation of the original vector $X$. The correctness of the above equation can be easily proofed algebraically. 

To reduce the calculation burden we can also slightly change equations $\hat{\Psi}_6(g_1)$ and $\hat{\Psi}_4(g_2)$ in Algorithm~\ref{alg:plugin}. It is easy to notice a symmetry, that is
\begin{equation}
K^{(6)} \left(\frac{X_i~-~X_j}{g_1} \right) 
= 
K^{(6)} \left(\frac{X_j~-~X_i}{g_1}\right). 
\end{equation}

So, the double summations 
can be changed and, consequently, the final formula for $\hat{\Psi}_6(g_1)$ has now the following form
\begin{equation}
\hat{\Psi}_6(g_1) 
\gets 
\frac{1}{n^2 g_1^7} \left[ 2 \left( \sum_{i=1}^{n}  \sum_{j=1, i<j}^{n} K^{(6)} \left( \frac{X_i - X_j}{g_1} \right) \right) + nK^{(6)}(0) \right]
\label{eq:PLUGIN-Psi6Estimate-2}
\end{equation}
(note for different summation ranges, the ,,2'' before sums and an extra factor added, that is $ nK^{(6)}(0)$).  
Obviously, the same concerns $K^{(4)}$ and $\hat{\Psi}_4(g_2)$
\begin{equation}
\hat{\Psi}_4(g_2) 
\gets 
\frac{1}{n^2 g_2^5} \left[ 2 \left( \sum_{i=1}^{n}\sum_{j=1, i<j}^{n} K^{(4)} \left( \frac{X_i - X_j}{g_2}\right)\right) + nK^{(4)}(0) \right].
\label{eq:PLUGIN-Psi4Estimate-2}
\end{equation}

Computation complexity of Steps IV and VI (double summations), where the symmetry property is used, still belongs to $O(n^2)$ complexity class
\begin{equation}
T(n) = \sum_{i=1}^{n}\sum_{j=i}^{n} T_k = \frac{1}{2} (n^2 + n ) T_k 
\end{equation}
where $T_k = T_1 + T_2 + T_3$, and $T_1$ represents computation time for the differences, $T_2$ represents division time, and $T_3$  represents time for computing  $K^{(6)}$ and $K^{(4)}$ polynomials.



\section{FPGA-based Implementation}\label{sec:fpga-impl}
\subsection{Xilinx's High Level Synthesis}

High Level Synthesis (HLS) is an automated design process that interprets an algorithmic description of a problem (given in high level languages C/C++) and translates this problem into a so called register-transfer level (RTL) HDL code. Then in turn this HDL code can be easily synthesized to the gate level by the use of a logic synthesis tool, like for example \emph{Xilinx ISE Design Suite}, \emph{Xilinx Vivado Design Suite}, \emph{Altera Quartus II}.

In this paper we discuss results obtained using a tool called \emph{Xilinx Vivado High Level Synthesis}, a feature of \emph{Vivado Design Suite}. This tool supports C/C++ inputs, and generates VHDL/Verilog/SystemC outputs. Other solutions are offered by \emph{Scala} programming language \cite{Bachrach2012} and a specialised high level synthesis language called \emph{Cx} \cite{Synflow}. It should also be mentioned that a similar tool called A++ is also available for Altera FPGA devices.



\subsection{Implementation Preliminaries}\label{sec:impl-prelim}
Before implementing the PLUGIN Algorithm \ref{alg:plugin} it is important to take some assumptions affecting both performance and resource consumption. 

\emph{The first assumption} is about a proper arithmetic used. The floating-point one gives very good range and precision. Unfortunately, from FPGA's point of view, this representation is very resource demanding. In contrast, the fixed-point arithmetic is much less resource demanding but its range and precision are more limited.



Hence, the exact fixed point representation was determined based on a careful analysis of the particular intermediate values taken during calculations. If the input dataset doesn't contain extremely large outliers (which suggests that such dataset should be first carefully analyzed before any statistical analysis taken) and if the \emph{z-score} standardization is used, $Q32.32$ fixed point representation is sufficient for all calculations (that is: integer part length $m=31$, fractional part length $n=32$, word length $N=64$ and the first bit represents the sign). Note also that as a result of the  \emph{z-score} standardization, the vales of $\hat{V}$, $\hat{\sigma}$, $\hat{\Psi}^{NS}_8$ are constant and this significantly simplifies the calculations.  
The fractional part does give the required precision. However, the integer part must also be sufficiently large, as $n^2$ factors are present in the PLUGIN algorithm. 

\emph{The second assumption} is about choosing the most adequate methods for calculating individual steps in Algorithm~\ref{alg:plugin}. Now it needs to be stressed that programming for FPGA devices differs considerably from programming for CPUs/GPUs devices. FPGA devices are built from a large number of simple logical blocks like: Look Up Tables (LUT), Flip-Flops (FF), Block RAMs (BRAM), specialized DSP units (DSP). These blocks can be connected each other and can implement only relatively low-level logical functions (the so called gates level). As a consequence, even very basic operations, like for examples the adder for adding two numbers must be implemented from scratch. In description of the PLUGIN Algorithm \ref{alg:plugin} one can easily indicate such operators like (a) addition, (b) subtraction, (c) multiplication, (d) division, (e) reciprocal, (f) exponent, (g) logarithm\footnote{Logarithm is not directly present in the PLUGIN mathematical formulas, but it is used while implementing higher order roots from the following definition $x^y = \exp(y \ln{x})$.}, (h) power, (i) square roots, (j) higher order roots. 


Our implementation utilizes the following methods: \emph{CORDIC} \cite{Cordic-1959},\cite{Cordic-1971} for calculating exponents and logarithms, divisions were replaced by multiplications and reciprocals, difference operators were replaced by addition of negative operands. 
Additionally, one extra implementation of the exponent function was used for calculations of $K^{(6)}$ and $K^{(4)}$ in Algorithm \ref{alg:plugin}. This implementation is based on the Remez algorithm \cite{Daili2013}, \cite{Rabiner1975}, \cite{Remez1934} and is open to 
pipelining. As a consequence, a significant speedup can be achieved during calculations of Steps IV and VI in Algorithm \ref{alg:plugin}.  

It is also worth to note that the authors' implementation of the division operator (base on multiplications and reciprocals; the reciprocal is based on the Newton method) is  significantly faster than the default division operator available in Vivado HLS.
Moreover, the another advantage of using our own operators, is that no IPCore (Xilinx's library of many specialized functions available for FPGA projects) is needed. As a consequence, the generated VHDL codes are more portable for FPGA chips from different than Xilinx vendors.


\emph{The third assumption} during implementing of the PLUGIN algorithm was to enable the nominal clock frequency of an FPGA chip used (see chapter \ref{sec:impl-results} for details).  
During experiments it was turned out that the usage of the original  division operator resulted in problems with reaching the required frequency. The authors' original implementation of the division operator (base on multiplications and reciprocals) solved this problem. 
	
\emph{The forth assumption} was that all the input datasets must be stored in the BRAM memory, which are available in almost all current FPGA chips. They have enough capacity to store truly large data, like even 500,000 elements or more.

\subsection{Implementation Details}\label{sec:impl-details}

\begin{figure}[!t]
	\centering
	\includegraphics[width=13cm]{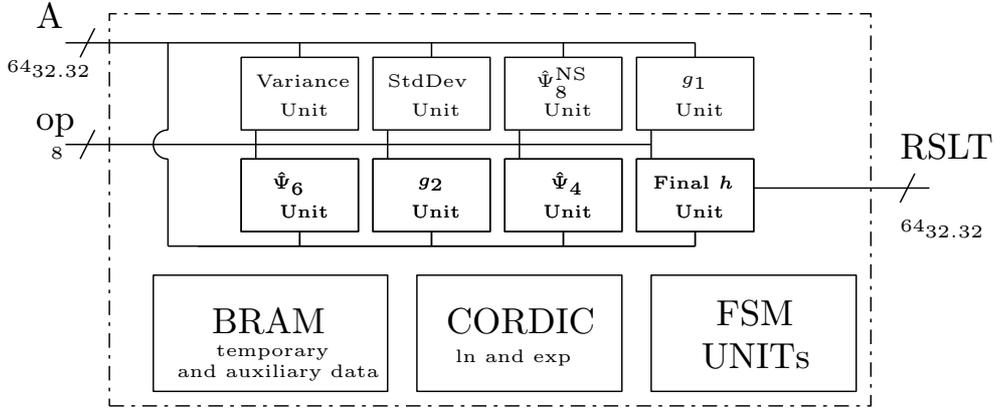}
	\caption{
		General overview of the main units for the FPGA-based PLUGIN algorithm implementation 
	} 
	\label{fig:plugin-unitis}
\end{figure}

In Figure~\ref{fig:plugin-unitis} we show the scheme of the PLUGIN implementation where all the main components are presented. They correspond literally to the seven steps shown in Algorithm~\ref{alg:plugin}. 

\begin{figure}[!t]
	\centering
	\includegraphics[width=13cm]{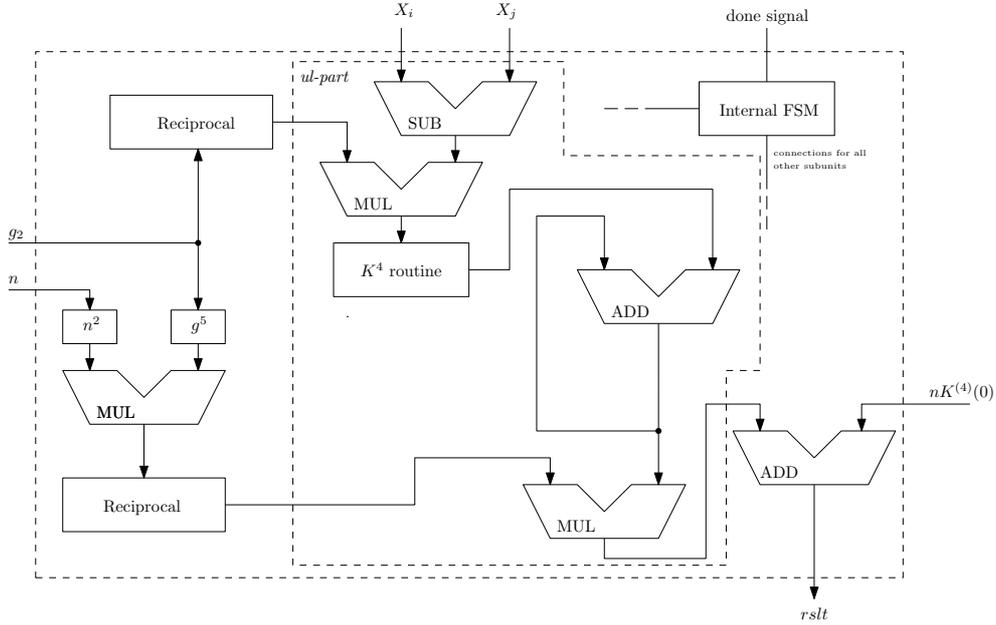}
	\caption{General architecture of the $\hat{\Psi}_4(g_2)$ unit at the block-level view. The extra frame called \emph{ul-part} shows the part of the Step VI in Algorithm~\ref{alg:plugin} where \emph{loop unrolling} can be used}
	\label{fig:one-block-architecture}
\end{figure}

Figure~\ref{fig:one-block-architecture} presents general architecture of the functional unit for computing  $\hat{\Psi}_4(g_2)$ (Step VI in Algorithm \ref{alg:plugin}). It is worth to note that the proper architecture of this unit must be reached during careful coding in Vivado HLS, using techniques like listed in section \ref{sec:impl-prelim}. 

We developed three different versions of the PLUGIN algorithm.

\emph{The first implementation}, called \emph{literal}, is just a literal rewriting of Algorithm~\ref{alg:plugin} (with the improvements (\ref{eq:PLUGIN-Psi6Estimate-2}) and (\ref{eq:PLUGIN-Psi4Estimate-2})). No additional actions were taken toward optimization of both execution time and resource requirements. This version can operate with any unscaled input data (assuming that all the inputs as well as all the internal results fulfil the fixed-point ranges that have been set). This version automatically (Vivado decides) utilizes pipelining. However, the pipelining doesn't make the implementation enough fast and additionally, large number of DSP blocks is used. FFs and LUTs usage is also quite big (see Table \ref{tab:resource-usage}).

\emph{The second implementation}, called \emph{minimal}, is written so that it is optimized for resource utilization, mainly the DSP units.  
To reduce the number of the DSP units some dedicated functions for addition and multiplication are required. Using Vivado HLS compiler's pragmas (\emph{\#pragma HLS INLINE off}) pipelining can be disabled (on default, during translation of the high level codes into HDL ones pipelining is enabled whenever it is possible). As can be observed in Table \ref{tab:resource-usage}, a significant reduction of the DSP units was achieved. It confirms the fact that Vivado HLS is very sensitive for the structure of the high level codes being translated into HDL ones. So that, to achieve good performance and resource usage many modifications of the high level codes are required.    

\emph{The third implementation}, called \emph{fast}, is written so that it is optimized for time execution. 
%
%
Addition and multiplication functions were implemented in two ways. In the first way (similar as in \emph{minimal} implementation) the pipelining is disabled, while in the second way it is enabled. The pipelined versions of the functions are used in Steps IV and VI in Algorithm \ref{alg:plugin} as these two steps are crucial for the final performance. Additionally, in these two steps a dedicated implementation of the exponent function was used (based on Remez algorithm which is more likely to pipelining). Also, a technique known as \emph{loop unrolling} was used in a manual manner (see sample codes in Figure \ref{fig:sample-loops}). Although Vivado HLS uses automatic loop unrolling, this feature doesn't work correctly in our algorithm (as it can operate with datasets of any size and the exact number of loops is not known in advance). 

The \emph{forth and fifth implementations} used during experiments (called \emph{CPU} and \emph{GPU} respectively) are the ones implemented and investigated in \cite{Gram-2013}. CPU implementation utilizes the SSE (Streaming SIMD Extensions) of the current multicore CPUs.

\subsection{Results}\label{sec:impl-results}

During all practical experiments the target \emph{Xilinx Virtex-7 xc7vx690tffg1761-2} device was used. Its nominal working frequency is 200MHz (or 5 ns for a single clock tact). CPU implementation was run on \emph{Intel Processor i7 4790k 4.0 GHz}. \emph{Geforce 480GTX} graphics card was used for GPU implementation. \emph{Vivado HLS ver. 2015.2} was used for developing all the FPGA implementations.


The summary of the \emph{resource consumption} is given in Table~\ref{tab:resource-usage}. Additionally, \emph{power consumption} is included. It is a real power (in Watts) taken by the FPGA chip after physical implementation of the PLUGIN algorithm using Vivado Design Suite. The power consumption of the FPGA implementations is significantly smaller comparing with the power consumption of the CPU and GPU implementations. 
The power consumption for the CPU and GPU used in our experiments are an average (catalogue-like) values. 
\begin{table}[!t]
	\centering
	\caption{Resources usage for three different FPGA implementations of the PLUGIN algorithms as well as CPU and GPU implementations. Additionally power consumption is included.} \label{tab:resource-usage}
	\begin{tabular}{cccccc}
		\hline
		\bf{Method} & \bf{BRAM 18k} & \bf{DSP} & \bf{FF} & \bf{LUT}  & \bf{Watts} \\ \hline
		literal  & 128 & 1164   & 80753  & 81995 & 3.938 \\ \hline
		minimal  & 128 & 240    & 15889  & 22895 & 1.153 \\ \hline
		fast     & 128 & 1880   & 85775 & 38050 & 6.963 \\ \hline
		CPU      & -   & -      & -      & -     & $\approx$ 88 \\ \hline
		GPU      & -   & -      & -      & -     & $\approx$ 250 \\ \hline				
	\end{tabular}
\end{table}

The summary of the \emph{execution times} for three different implementations of the PLUGIN algorithm, as well as CPU and GPU ones is given in Table~\ref{tab:execution-times}. 
The \emph{minimal} and the \emph{fast} implementations were run on 200MHz nominal clock while the \emph{literal} implementation was run with 166 MHz nominal clock. This frequency degradation was caused mainly because of some limitations of the original division operator implemented in Vivado HLS. 
\begin{table}[!t]
	\centering
	\caption{Execution times (in sec.) for three different FPGA implementations of the PLUGIN algorithm and for CPU and GPU implementations} \label{tab:execution-times}
	\begin{tabular}{cccccc}
		\hline
		\bf{n} & \bf{literal} & \bf{minimal} & \bf{fast} & \bf{CPU} & \bfseries{GPU} \\ \hline
  128 & 0.0555 & 0.0324 & 0.000276 & 0.0210 & 0.00699  \\ \hline
  256 & 0.2266 & 0.1363 & 0.000560 & 0.0252 & 0.00788  \\ \hline
  384 & 0.5155 & 0.3152 & 0.000889 & 0.0322 & 0.00947  \\ \hline
  512 & 0.9112 & 0.5513 & 0.001257 & 0.0346 & 0.00962  \\ \hline
  640 & 1.4466 & 0.8968 & 0.001667 & 0.0361 & 0.01063  \\ \hline
  768 & 2.1023 & 1.3205 & 0.002114 & 0.0375 & 0.01172  \\ \hline
  896 & 2.8771 & 1.8232 & 0.002606 & 0.0405 & 0.01447  \\ \hline
 1024 & 3.7666 & 2.3926 & 0.003140 & 0.0427 & 0.01641  \\ \hline
	\end{tabular}
\end{table}

Of course the best performance was achieved for the \emph{fast} implementation (even compared to the \emph{CPU} and to the \emph{GPU} implementations). This is the result of combination of the following three optimization techniques used: (a) implementation of some dedicated arithmetic operators, (b) a proper exponential function approximation and (c) the \emph{for} loops unrolling. 

A very significant speedup was achieved comparing the \emph{fast} and the \emph{literal} implementation (average speedup about 760, see Table \ref{tab:speedups}). The \emph{fast} implementation is faster then the \emph{CPU} implementation (average speedup about 32, see Table \ref{tab:speedups}). The \emph{fast}  implementation is also faster then the \emph{GPU} implementation (average speedup about 10, see Table \ref{tab:speedups}). 
\begin{table*}[!t]
	\centering
	\caption{Speedups for three different FPGA implementations of the PLUGIN algorithm and for CPU and GPU implementations } \label{tab:speedups}
	\begin{tabular}{ccccc}
		\hline
		\bf{n} & \bf{literal/fast} & \bf{minimal/fast} & \bf{CPU/fast} & \bf{GPU/fast}  \\ \hline
  128 &  201 & 118 & 76 & 25  \\ \hline
  256 &  404 & 243 & 45 & 14  \\ \hline
  384 &  580 & 354 & 36 & 11  \\ \hline
  512 &  725 & 439 & 28 &  8  \\ \hline
  640 &  868 & 538 & 22 &  6  \\ \hline
  768 &  994 & 625 & 18 &  6  \\ \hline
  896 & 1104 & 700 & 16 &  6  \\ \hline
 1024 & 1200 & 762 & 14 &  5  \\ \hline 
 	\end{tabular}
\end{table*}


The summary of the \emph{accuracy} for three different implementations of the PLUGIN algorithm is given in Table~\ref{tab:accuracy}. $h_{ref}$ is the reference bandwidth calculated in double floating point arithmetic (in C++ program, 15--17 significant decimal digits). It is worth to note that the relative errors for \emph{literal}, \emph{minimal} and \emph{fast} implementations are very small (not more than $0.004\%$). In practical applications such small values can be in fact neglected. 
\begin{table}[!t]
	\centering
	\caption{Accuracy (relative error) for three different FPGA implementations of the PLUGIN algorithms.  $h_{ref}$ was calculated in C++ direct implementation of Algorithm \ref{alg:plugin} in floating point double arithmetic (15--17 significant decimal digits). 
	$|\delta_x| = \frac{|h_{method} - h_{ref}|} {|h_{ref}|}*100\%$ where $h_{method}$ is $h_{literal}$,  $h_{minimal}$ or $h_{fast}$} \label{tab:accuracy}
	\begin{tabular}{cccc}
		\hline
	    \bf{n} & $h_{literal}$ & $h_{ref}$ & \bfseries{\shortstack{$|\delta_x|$ $(\%)$}} \\ \hline
		128 & 0.304902711650357 & 0.304902701728222 & 3.25e-06  \\ \hline
		256 & 0.227651247521862 & 0.227651285449348 & 1.67e-05  \\ \hline
		384 & 0.202433198224753 & 0.202433187549741 & 5.27e-06  \\ \hline
		512 & 0.242707096505910 & 0.242707026022425 & 2.9e-05  \\ \hline
		640 & 0.190442902734503 & 0.190443702342891 & 0.00042  \\ \hline
		768 & 0.175199386896566 & 0.175199406819444 & 1.14e-05  \\ \hline
		896 & 0.172251554206014 & 0.172251524317464 & 1.74e-05  \\ \hline
	   1024 & 0.174044180661440 & 0.174044236921001 & 3.23e-05  \\ \hline

	    \bf{n} & $h_{minimal}$ & $h_{ref}$ &  \bfseries{\shortstack{$|\delta_x|$ $(\%)$}} \\ \hline
		128 & 0.304902980336919 & 0.304902701728222 & 9.14e-05  \\ \hline
		256 & 0.227651586290449 & 0.227651285449348 & 0.000132  \\ \hline
		384 & 0.202433346537873 & 0.202433187549741 & 7.85e-05  \\ \hline
		512 & 0.242707266006619 & 0.242707026022425 & 9.89e-05  \\ \hline
		640 & 0.190443017752841 & 0.190443702342891 & 0.000359  \\ \hline
		768 & 0.175199396442622 & 0.175199406819444 & 5.92e-06  \\ \hline
		896 & 0.172251742798835 & 0.172251524317464 & 0.000127  \\ \hline
	   1024 & 0.174044403014705 & 0.174044236921001 & 9.54e-05  \\
        \hline			   
	    \bf{n} & $h_{fast}$ & $h_{ref}$ &  \bfseries{\shortstack{$|\delta_x|$ $(\%)$}} \\ \hline 
        128 & 0.304901758907363 & 0.304902701728222 & 0.000309 \\ \hline
        256 & 0.227651913650334 & 0.227651285449348 & 0.000276 \\ \hline
        384 & 0.202433891594410 & 0.202433187549741 & 0.000348 \\ \hline
        512 & 0.242707268567756 & 0.242707026022425 & 9.99e-05 \\ \hline
        640 & 0.190443484811112 & 0.190443702342891 & 0.000114 \\ \hline
        768 & 0.175199736841023 & 0.175199406819444 & 0.000188 \\ \hline
        896 & 0.172251721611246 & 0.172251524317464 & 0.000115 \\ \hline
       1024 & 0.174044031649828 & 0.174044236921001 & 0.000118 \\ \hline	      
	\end{tabular}	   
\end{table}

\begin{figure}[!t]
	\centering
	\includegraphics[width=12cm]{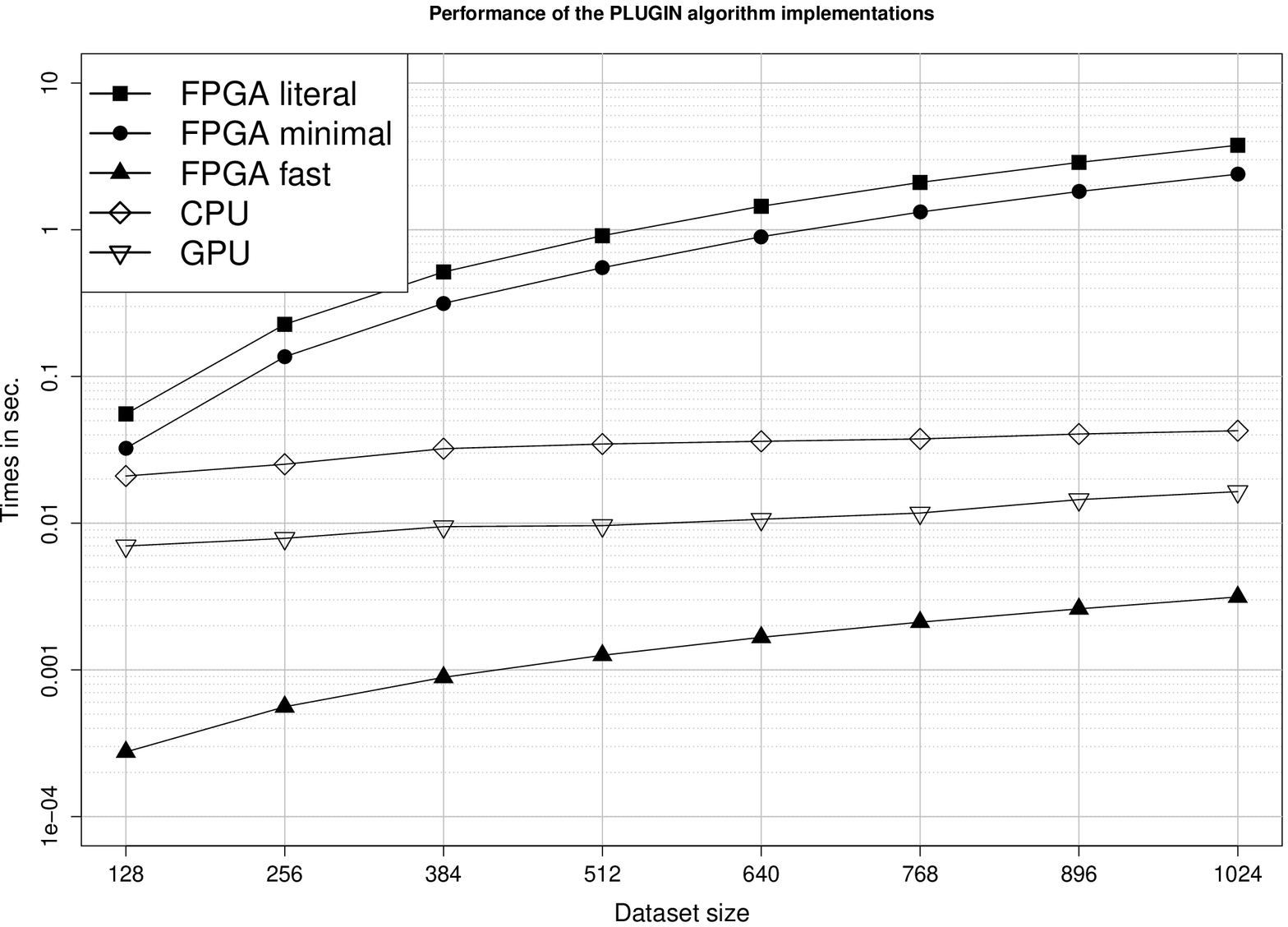}
	\caption{Performance and scalability of different PLUGIN algorithm implementations (for better readibility log scale for $Y$ axis is used)} 
	\label{fig:fpga-implementation-speedup}
\end{figure}

The summary of the \emph{scalability} of different PLUGIN algorithm implementations is presented in Figure \ref{fig:fpga-implementation-speedup}. Scalability of the FPGA implementations is nearly linear, which is a very welcome behavior. The corresponding results for \emph{CPU} and \emph{GPU} implementations can be found in \cite{Gram-2013}. The figure is in fact a graphical summary of data given in Table \ref{tab:execution-times}. 



Simplified source codes of the three FPGA implementations are presented in Figure \ref{fig:sample-loops}. Complete source codes (C++ and resulted Vivado HLS translations into VHDL) are available on request. The first version is just the literal implementation of the step VI in Algorithm \ref{alg:plugin} in C language. Unfortunately, as can be observed in Table \ref{tab:execution-times} and in Figure \ref{fig:fpga-implementation-speedup} such implementation is very slow. In the second version multiplications and additions are realized using dedicated functions (\emph{fADD}, \emph{fMUL}). Also a dedicated function for reciprocal operator was implemented. In the third version much more modification was implemented. First, loop unrolling was used, second, Vivado HLS pragmas were used and third, multiplications and additions were realized using dedicated functions with pipelining enabled (\emph{pfADD}, \emph{pfMUL}).   

\begin{figure}
\begin{footnotesize}
\begin{verbatim}
// literal implementation
psi4_f1: for( i=0; i<N; i++ ) {
    psi4_f2: for( j=i+1; j<N; j++ ) {
        s = s + k4( ( ( x[i] - x[j] ) / g2) );
    }
}

// minimal implementation
rg2 = reciprocal( g2 );
psi4_f1: for( i=0; i<N; i++ ) {
    psi4_f2: for( j=i+1; j<N; j++ ) {
        s = fADD( s, k4( fMUL( fADD( x[i], -x[j] ), rg2 ) ) );
    }
}

// fast implementation
rg2 = reciprocal( g2 );
psi4_f1: for( i=0; i<N; i++ ) {
    psi4_f2: for( j=i+1; j<N; j+=2 ) {
        
        #pragma HLS EXPRESSION_BALANCE
        #pragma HLS PIPELINE
        
        if( j == i+1 ) tmp = 0.0;
        if( j<N ) { tmp1 = 0.0; tmp2 = 0.0; }
        psi4_f1_b0: {
            tmp1a = pfADD( x[i], -x[j] );
            tmpva = pfMUL( tmp1a, rg2 );
            tmp1 = k4( tmpva );
        }
        psi4_f1_b1: {
            if( (j+1) < N ) {
                tmp1b = pfADD( x[i], -x[j+1] );
                tmpvb = pfMUL( tmp1b, rg2 );
                tmp2 = k4( tmpvb );
            }
        }
        if( j<N ) { tmp = pfADD( tmp, tmp1 ); tmp = pfADD( tmp, tmp2 ); }
        if( j+2>=N ) s = pfADD (s, tmp );
    }
}
\end{verbatim}
\end{footnotesize}
\caption{Three fundamental methods of the \emph{for} loop implementation used in $\hat{\Psi}_4(g_2)$ calculation (step VI in Algorithm \ref{alg:plugin}, step IV is implemnented in the same way). In the \emph{fast} implementation the \emph{loop unrolling} is used twice. fADD, fMUL functions don't utilize pipelining, while pfADD i pfMUL functions do it}
	\label{fig:sample-loops}
\end{figure}

\clearpage
\section{Conclusions}\label{sec:concluding}

HLS tools are competitive with manual design techniques using HDLs. Implementation time of complex numerical algorithms can be essentially reduced (comparing to direct coding in HDL languages). 

Unfortunately, to obtain efficient FPGA implementations, many changes to source codes are required, comparing to equivalent implementations for CPUs and/or GPUs. This is because FPGA devices use specific primitives like DSP, BRAM, FF and LUT blocks and programmers should control their utilization manually. However, this control is performed on the level of C/C++ codes, not the HDL ones. It is also worth to stress that using the HLS approach allows to obtain implementations which are often faster than CPU and/or GPU counterparts.

Another crucial motivation for replacing GPU or CPU solutions by their FPGA equivalents is power consumption. FPGA can settle for single Watts, while CPU or GPU counterparts typically take tens/hundreds of Watts or even more.

Another possible step toward fast implementations of numerical algorithms could be considering of using modern DSP chips which offer many interesting possibilities.  	

\subsection*{Acknowledgements}

We would like to thank for useful discussions with colleagues at the Institute of Control and Computation Engineering (ISSI) of the University of Zielona G\'ora, Poland. We would like also to thank to anonymous referees for useful comments on the preliminary version of this paper. The numerical results were done using the hardware and software available at the ''FPGA/GPU $\mu$-Lab'' located at the Institute of Control and Computation Engineering of the University of Zielona G\'ora, Poland.

\end{document}